\documentclass{article}
\usepackage{epsfig}  

 \normalmarginpar
\twocolumn
 \makeatletter
 \gdef\@proofbox{\relax}
 \long\def\proofbox#1{\gdef\@proofbox{#1}}
 \proofbox{\small{\sl PhysComp96}\\\ifx\UNDEF\@fullpaper
     Extended abstract\else Full paper\fi}

 \gdef\fullpaper{\gdef\@fullpaper{}}

 \def\affil#1{\\{\small#1\par}}
 \gdef\@author{John Doe1\affil{No-Name University, Shipping Dept.}}
 \long\def\author#1{\gdef\@author{#1}}

 \gdef\@abstract{}
 \long\def\abstract#1{\gdef\@abstract{#1}}

\def\@maketitle{\newpage\leavevmode
  \begin{minipage}[t]{0.30\textwidth}
    \hrule height0pt
    \raggedright
    \mbox{}\par
    \@proofbox
  \end{minipage}\relax
  \begin{minipage}[t]{0.70\textwidth}
    \hrule height0pt
    \raggedleft
    \LARGE\@title\par
    \vskip4pt
    \large\@author
  \end{minipage}
  \vskip8pt
  \ifx\@abstract\@empty\else{\vskip.5em\leftskip1.5in\parskip4pt\small\@abstract\par\vskip.5em}\fi
  \rule{\textwidth}{0.4pt}
  \vskip16pt}

\DeclareRobustCommand\em
        {\@nomath\em \ifdim \fontdimen\@ne\font >\z@
                       \upshape \else \slshape \fi}

\def\@begintheorem#1#2{\sl \trivlist \item[\hskip \labelsep{\bf #1\ #2}]}
\def\@opargbegintheorem#1#2#3{\sl \trivlist
     \item[\hskip \labelsep{\bf #1\ #2\ (#3)}]}

 \newcommand{\eq}[1]{(\ref{eq.#1})}	
 \newcommand{\fig}[1]{Fig.~\ref{fig.#1}}

 \newcommand{\eqlabel}[1]{\label{eq.#1}}

  \def\@arabic#1{\number #1} 

\long\def\@makecaption#1#2{
	\vskip\abovecaptionskip
	\sbox\@tempboxa{{\small #1: #2}}%
	\ifdim\wd\@tempboxa>\hsize
	    {\small #1: #2\par}
	\else
	   \global\@minipagefalse
	   \hbox to\hsize{\hfil\box\@tempboxa\hfil}
	\fi
	\vskip \belowcaptionskip}

\def\figstrut#1{\hbox to\linewidth{\vrule height#1\hfill}}

\newcommand{\figonecol}[3]{
\begin{figure}[!htb]
 \centering\leavevmode#2%
 \caption{#3}
 \label{fig.#1}
\end{figure}                 }

\newcommand{\figtwocol}[3]{
\begin{figure*}[!t]
 \centering\leavevmode#2%
 \caption{#3}
 \label{fig.#1}
\end{figure*}                 }

\def\comppad{\thinspace}
\def\comp{\comppad\begingroup \tt \let\do\@makeother \dospecials 
          \@ifstar{\@scomp}{\@comp}}
\def\@scomp#1{\def\@tempa ##1#1{##1\endgroup\comppad}\@tempa}
\def\@comp{\obeyspaces \frenchspacing \@scomp}

\makeatother

 \fullpaper
 \title{A Framework for Quantum Search Heuristics}
 \author{Tad Hogg
	\affil{Xerox Palo Alto Research Center\\
	3333 Coyote Hill Road\\
	Palo Alto, CA 94304\\
	hogg@parc.xerox.com}}
 \date{11 April 1996}
 \abstract{
A quantum algorithm for combinatorial search is
presented that provides a simple framework for utilizing search
heuristics. The algorithm is evaluated in a new case that is an unstructured
version of the graph coloring problem. It performs significantly better than the
direct use of quantum parallelism, on average, in cases corresponding to
previously identified {``}phase transitions{''} in search
difficulty. The conditions underlying this improvement are described.
Much of the algorithm is independent of particular problem instances,
making it suitable for implementation as a special purpose device.
}

\begin{document}

\maketitle

\section{Introduction}
Quantum computers~\cite{benioff82,bernstein92,deutsch85,deutsch89,feynman86,lloyd93,divincenzo95}
use {\em quantum parallelism}, i.e., the ability to
operate simultaneously on a superposition of many classical states, and
{\em interference} among different
computational paths. A measurement on a superposition gives a definite
result, with probabilities determined by the amplitudes of the
superposition. A successful algorithm is one that uses these
capabilities to arrange for a large probability of a desired result,
e.g., a solution to a search problem.

However, there are two major difficulties with quantum computers. 
First, they are difficult to implement~\cite{landauer94a}. 
Second, the physical
restriction to unitary linear operations makes quantum computers
difficult program effectively. An encouraging development with respect
to the design of algorithms is a method for efficiently factoring
integers~\cite{shor94}, a problem that appears to
be intractable for classical computers. Since this relies on specific
properties of the factoring problem, the extent to which effective
quantum algorithms can be developed for general combinatorial search
remains an open question.

To help address this question empirically, it would be useful to
have a simple framework for exploring the use of search heuristics with
quantum algorithms. Specifically, such a framework should allow a
variety of heuristic search methods to be programmed while automatically
preserving a number of desirable properties that simplify potential
hardware implementations. This could serve to bridge the gap between
discussions of hardware, with the focus usually on bit-level operations,
and abstract theoretical discussions, with the focus on Turing machines
and general unitary operations.

There are a number of desirable properties such a framework should
provide for a general quantum search algorithm. First, the number of
computational steps required should be determined a priori. This avoids the
question of when to make the final measurement, and could limit the
difficulties of decoherence compared to running a program where the
number of steps required can vary greatly from one problem instance to
the next. Second, it is useful if much of the complexity of the
algorithm can be made independent of the details of individual
instances. This can facilitate implemention with a
special purpose search device. Third, the framework should easily allow
applying heuristics that incorporate additional knowledge about the
structure of particular search problems, in a manner analogous to the
heuristics used to dramatically improve many classical search
strategies.

Another property is given by recent studies of the nature of
classes of combinatorial search problems and relates to how methods
are evaluated. Most theoretical analyses of search algorithms focus on worst
case behavior. However, in practice it is often more important to examine the
behavior of algorithms for typical or average search problems. Thus even
if quantum computers are not applicable to {\em all}
combinatorial search problems, they may still be useful for many
instances encountered in practice. This is an important distinction
since typical instances of search problems are often much easier to
solve than is suggested by worst case analyses, though even typical costs
often grow exponentially on classical machines.

In fact, the hard
instances are not only rare but also concentrated near abrupt
transitions in problem behavior analogous to physical phase
transitions~\cite{cheeseman92,hogg94b,web.hogg94}.
This result applies to many classical search methods that use problem
structure to guide choices, but not to generate-and-test, where possible
solutions are examined sequentially. Similarly, the most direct use of
quantum parallelism for search, in which all states are simultaneously
checked for consistency and then a measurement made, is equivalent to
generate-and-test and does not exhibit the transition behavior. 
This limitation may apply to any search algorithm that uses
quantum parallelism without also using interference~\cite{cerny93}. Thus a
check on whether a quantum algorithm is fully exploiting problem
structure, through interference among different computational paths, is
whether it exhibits the transition behavior. This is another useful
property to incorporate in a general search framework.

In this paper, we briefly describe a previously proposed general search
mechanism and discuss how it meets these criteria as well as some
implementation issues. To examine its
generality, its behavior is then evaluated empirically for a new problem
ensemble, motivated by the NP-complete graph coloring problem. As with
other previously studied problems, this shows a significant enhancement
of probably to obtain a solution compared to the direct use of quantum
parallelism and the transition behavior. We give a more detailed look at
the underlying basis for this improvement to see how heuristics might be
usefully applied. Finally, some open issues are presented.

\section{A Quantum Search}
Combinatorial search can be viewed as finding, from among
$N$ given items, a set of size
$L$ satisfying specified constraints.
Such a set is a solution to the problem. The constraints, in turn,
can be specified by {\em nogoods}, sets
of items that are inconsistent. Because the supersets of a nogood are
also nogood, sets of items can be viewed as forming a lattice structure
consisting of levels from 0 to $N$.
Level $i$ in this lattice contains all
sets of size $i$, and these are linked
to their supersets at the next level and subsets at the previous one.
This lattice, describing the consistency relationships among sets, is
the {\em deep structure} of the
combinatorial search problem~\cite{williams92}. This structure for
$N=4$ is shown in \fig{lattice}.
Solutions are found among the sets at level
$L$. Notationally, we denote the size of a set $s$ by $|s|$.

\figonecol{lattice}{
\epsfig{file=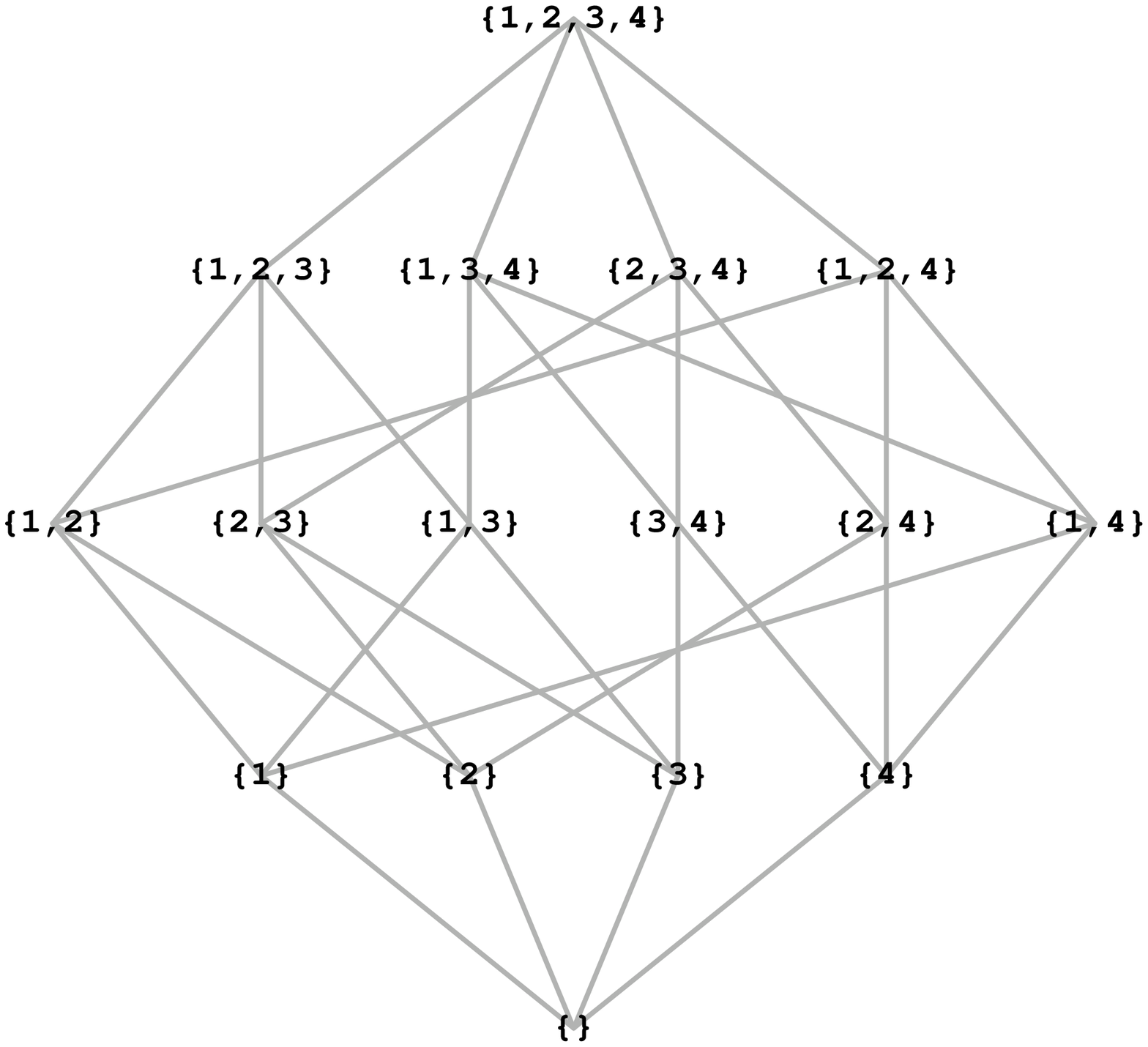,width=3.5in}
}{Structure of the set lattice for a problem with four items. The subsets of 
\({\left\lbrace 1,2,3,4\right\rbrace }\) are grouped into levels by size and 
lines drawn between each set and its immediate supersets and subsets. 
The bottom of the lattice, level 0, represents the single set of size zero, 
the four points at level 1 represent the four singleton subsets, etc.}

This abstract description of search is less commonly used than
other representations, which are more compact and efficient
for classical search algorithms. It is introduced
here as a useful basis for quantum searches and because it applies to
many search problems. These include constraint satisfaction problems
(CSPs)~\cite{mackworth92} such as graph coloring
and satisfiability. For example, in coloring an
$n$-node graph with
$c$ colors an item is a pair
\(
{\left( \nu ,\kappa \right) }\) consisting of a node $\nu$ in the graph and a color
$\kappa$ for it. Thus there are \(
N=nc\) items and a solution consists of \(
L=n\) such items that give a unique color to each node and
distinct colors to each pair of nodes in the graph that are linked by an
edge. Each edge is a constraint which gives
$c$ nogoods, each consisting of a pair
of items with the same color for both of the nodes linked by that edge.
This search problem is known to be NP-complete for a fixed
$c$ (at least equal to 3) as
$n$ grows. From this example, we see
that an interesting scaling regime for combinatorial search is for the
nogoods of the constraints to have a fixed, small size (e.g., at level 2
in the lattice for graph coloring) while the number of items and the
size of solutions grows linearly with problem size.

There are many paths through the lattice from small sets to the
larger sets at the solution level. For example, if \(
N=4\) and \(
L=2\), there are two paths from level zero to each set at
level 2 in the lattice. E.g., the set \(
{\left\lbrace 1,2\right\rbrace }\) is obtained via the paths \(
\emptyset {\rightarrow}{\left\lbrace 1\right\rbrace }{\rightarrow}{\left%
\lbrace 
1,2\right\rbrace }\) and \(
\emptyset {\rightarrow}{\left\lbrace 2\right\rbrace }{\rightarrow}{\left%
\lbrace 
1,2\right\rbrace }\). These multiple paths can be used to create
interference with a quantum computer operating on a superposition of
sets. One way to do this uses the fact that for search problems in NP 
it is relatively easy to
test whether a particular set satisfies the given constraints. Thus, we
can adjust the phase of the amplitudes along each path based on whether
the associated set is consistent with the constraints. These phase
adjustments attempt to create destructive interference among paths
leading to nonsolutions and constructive interference for solutions. To
the extent this is successful, amplitude will be concentrated into
solution states, giving a relatively high probability to obtain a
solution in the final measurement. A quantum computer with
$N$ bits can simultaneously operate on
superpositions of all \(
2^{N}\) subsets of \(
{\left\lbrace 1,\ldots ,N\right\rbrace }\). Thus we can start with a 
superposition consisting of
small subsets, and move up one level at a time to the solution level,
exploring all possible paths

Following this general concept, there are many possible mappings
among these sets. A particularly simple method is to divide the mapping
into two parts. The first is problem-independent and moves amplitude
from sets of a given size to the next larger size. The second is an
adjustment to the phases of the sets at the new level. A
simple choice for the phase adjustment process is to change the sign of
the amplitude of each inconsistent set encountered. While this is not
always the optimal phase policy, it has the advantage of being easily
computable since it operates independently on each set and, as shown
below, is quite effective.

The mapping from one level to the next is a more complex quantum
operation as it involves mixing the amplitude from different states. On
the other hand, this mapping is independent of the details of individual
problem instances which could simplify its implementation as a special
purpose search device instead of a program operating on a universal
quantum computer. It is this part of the algorithm that gives rise to
interference effects by mixing the contribution from different paths
through the lattice. The operation used here is motivated by
breadth-first classical search where each consistent set is extended to
each possible superset at the next level. Such a mapping is not unitary,
though it is nearly so. Thus a simple choice for the quantum mapping is
to use that unitary operator that is closest~\cite{golub83} to the 
mapping of sets equally to their
supersets. The resulting matrix elements have a simple structure,
depending only on the size of the intersection of the corresponding
sets~\cite{hogg95d}.

\subsection{The Search Method}
The search algorithm starts by evenly dividing
amplitude among the goods at a low level of the lattice. For problems
such as graph coloring where the constraint nogoods involve only two
items at a time, a reasonable choice is to start at level 2, where the
number of sets is proportional to \(
N^{2}\). Then for each level from 2 to \(
L-1\), we adjust the phases of the states depending on
whether they are  consistent and map to the next level. Let
\(
\psi ^{{\left( j\right) }}_{s
}\) be the amplitude of the set
$s$ at level
$j$ after completing step $j-2$ of the algorithm. The initial condition is
\(
\psi ^{{\left( 2\right) }}_{s
}=1/\sqrt{N_{\rm goods}} \) if the set $s$ at
level 2 is good, i.e., consistent with the constraints, and otherwise
\(
\psi ^{{\left( 2\right) }}_{s
}=0\). Here \(
N_{\rm goods}\) is the number of consistent sets at level 2.

Each step of the algorithm moves up one level giving
\begin{equation}\eqlabel{map}\vcenter{\halign{\strut\hfil#\hfil&#\hfil\cr 
$\displaystyle{\psi ^{{\left( j+1\right) }}_{%
r}=\sum _{k=0}^{j}a^{{\left( 
j\right) }}_{k}\sum _{{\left| 
r\cap s\right| }=k}\rho _{s}\psi ^{%
{\left( j\right) }}_{s}}$\cr 
}}\end{equation}
where \(
a^{{\left( j\right) }}_{k}\) is the problem-independent matrix
element\footnote{These values are available on the World Wide
Web~\cite[online appendix]{hogg95d}.} mapping from sets
$s$ of size
$j$ to those
$r$ of size \(
j+1\) that have $k$
elements in common with $s$,
\(
\rho _{s}\) is the phase assigned to the set
$s$ after testing whether it is nogood,
and the inner sum is over all sets
$s$ of size $j$ that have
$k$ items in common with
$r$. That is, \(
\rho _{s}=1\) when $s$ is a good, and otherwise \(
\rho _{s}=-1\).

After $L-2$ steps we measure the state, obtaining a single
set. This set will be a solution with probability
\begin{equation}\eqlabel{p(soln)}\vcenter{\halign{\strut\hfil#\hfil&#\hfil\cr 
$\displaystyle{P_{\rm soln}=\sum _{s}p{\left( 
s\right) }}$\cr 
}}\end{equation}
with the sum over solution sets and \(
p{\left( s\right) }={\left| \psi ^{%
{\left( L\right) }}_{s}\right| }^{%
2}\) is the probability to obtain the set
$s$ with the measurement of the final
state.

\subsection{Classical Simulation}
The study of the average or typical behavior of search heuristics
relies primarily on empirical evaluation. This is because the
complicated conditional dependencies in search choices made by the
heuristic often preclude a simple theoretical analysis, although
phenomenological theories can give an approximate description of some
generic behaviors~\cite{hogg94b,kirkpatrick94,monasson96}. Thus as a
practical matter, the search framework described here should allow for
empirical evaluation of various heuristic methods on existing classical
computers.

Unfortunately, the exponential slowdown and growth in memory
required for such a simulation severely limits the feasible size of the
search problems. To see this, note that
\eq{map} consists of a matrix
multiplication on a vector of size \(
N_{j}={N \choose j}\), equal to the number of sets at level
$j$ of the lattice, to produce a new
vector of size \(
N_{j+1}\). A direct evaluation of this mapping requires of
order \(
N_{j}N_{j+1}\) multiplications. To reach the solution level at
\(
L=O{\left( N\right) }\) requires mapping at relatively high levels in the
lattice where \(
j=\eta N\) with $\eta$ constant. In this case \(
\ln N_{j}\sim Nh{\left( \eta \right) }
\) where \(
h{\left( \eta \right) }=-\eta \ln 
\eta -{\left( 1-\eta \right) }\ln 
{\left( 1-\eta \right) }\), so the classical computation cost scales according
to \(
e^{2h{\left( \eta \right) }N}\), which is \(
2^{2N}\) when \(
\eta =1/2\).

The cost of the classical simulation can be reduced substantially
(though still growing exponentially) by exploiting the map{'}s simple
structure with a recursive evaluation. This is done by expanding the
map in \eq{map} to include {\em all} \(
2^{N}\) sets of $N$ items
and then writing \eq{map} as 
\begin{equation}\vcenter{\halign{\strut\hfil#\hfil&#\hfil\cr 
$\displaystyle{\psi ^{{\left( j+1\right) }}_{%
r}}$\hfilneg&$\displaystyle{{}=\sum _{k=0}^{j}
a^{{\left( j\right) }}_{k}V_{%
rjk}}$\cr 
$\displaystyle{V_{rjk}}$\hfilneg&$\displaystyle{{}\equiv \sum _{%
 {{\left| r\cap s\right| }=k} \atop {%
{\left| s\right| }=j}}x_{%
s}}$\cr 
}}\end{equation}
where \(
x_{s}\equiv \rho _{s}\psi ^{{\left( 
j\right) }}_{s}\) is the original state modified by the choice of
phases for each set, which is nonzero only for sets of size
$j$.

From the definition of \(
V_{rjk}\), it follows that 
\begin{equation}\eqlabel{V relation}
\vcenter{\halign{\strut\hfil#\hfil&#\hfil\cr 
$\displaystyle{\sum _{ {r^{\prime}\subset r} \atop
{{\left| r^{\prime}\right| }={\left| 
r\right| }-1}}V_{r^{\prime
}jk}={\left( {\left| r\right| }
-k\right) }V_{rjk}+{\left( k+1\right) }
V_{rj,k+1}}$\cr 
}}\end{equation}
The sum in this expression is only over the \(
{\left| r\right| }\) subsets of $r$ of
size \(
{\left| r\right| }-1\) and so can be rapidly computed for a given set
$r$. From this relation,
\(
V_{rj,k+1}\) can be determined easily given \(
V_{rjk}\) and the values for the subsets of
$r$ with one element removed. Hence, from
the values for \(
V_{rj0}\) we can readily determine the
\(
V_{rjk}\) for \(
k>0\) iteratively starting from \(
r=\emptyset \) at the bottom of the lattice and moving up a level at
a time until the values at level \(
j+1\) are determined.

To evaluate \(
V_{rj0}\) define the matrix
$A$, of size \(
2^{N}\times 2^{N}\), with elements \(
A_{rs}\) equal to zero if \(
{\left| r\cap s\right| }>0\), i.e., the sets have an element in common, and one
otherwise. Let \(
y_{r}=\sum _{s}A_{rs}x_{%
s}\). Because \(
x_{s}\) is zero unless the set
$s$ is of size
$j$, we have
\begin{equation}\eqlabel{Vrj0}
V_{rj0}=y_{r}
\end{equation}

The product \(
A{\bf x}\) can be computed recursively. To see this consider the
sets ordered by the value of the integer with corresponding binary
representation, e.g., the sets without item
$N$ come before those with
$N$. For example, the sets for
\(
N=3\) are ordered as $\{\}$, $\{1\}$, $\{2\}$, $\{1,2\}$, $\{3\}$, 
$\{1,3\}$, $\{2,3\}$
and $\{1,2,3\}$. In this ordering, the matrix
$A$ has the recursive decomposition
\begin{equation}\vcenter{\halign{\strut\hfil#\hfil&#\hfil\cr 
$\displaystyle{A={\left( 
\matrix{A^{\prime}&A^{\prime}\cr 
A^{\prime}&0\cr 
}\right) }}$\cr 
}}\end{equation}
where \(
A^{\prime}\) is the same matrix but defined on subsets of
\(
{\left\lbrace 1,\ldots ,N-1\right\rbrace }\) and 0 is the matrix all 
of whose entries equal zero.
We can then compute
\begin{equation}\vcenter{\halign{\strut\hfil#\hfil&#\hfil\cr 
$\displaystyle{A{\bf x}={\left(  {A^{\prime}{\bf x^{%
{\left( 1\right) }}}+A^{\prime}
{\bf x^{{\left( 2\right) }}}} \atop
{A^{\prime}{\bf x^{{\left( 1\right) }
}}}\right) }}$\cr 
}}\end{equation}
where \(
{\bf x^{{\left( 1\right) }}}\) and \(
{\bf x^{{\left( 2\right) }}}\) denote, respectively, the first and 
second halves of
the vector {\bf x} (i.e., corresponding to sets
without $N$ and with
$N$ respectively). Thus the cost to
compute \(
A{\bf x}\) is \(
C{\left( N\right) }=2C{\left( N-1\right) }
+O{\left( 2^{N}\right) }\) resulting in an overall cost of order
\(
N2^{N}\). While still exponential, this improves substantially
on the cost for the direct evaluation at high levels of the
lattice.

This calculation can be improved further in two ways. First, when
computing {\bf y} recursively store only the
components for sets of size \(
j+1\) or smaller. The values for larger sets are never
used. Second,  there is no need to explicitly compute and store the
\(
V_{rjk}\) values individually. This is because we are only
interested in a particular linear combination of these values in
\eq{map} for sets $r$ of size
\(
j+1\). Using \eq{V relation}
this can be determined from a linear combination of values for sets of
size $j$, and so on down to the bottom
of the lattice. The net result of this process means the map of
\eq{map} becomes a product of matrices,
one for each step up the lattice, which in turn multiplies the vector
\(
A{\bf x}\) computed recursively as described above.

The basic components of this computation, a recursive matrix
combined with a simple relation to build up the required linear
combinations, are not unique. Other possibilities are useful to keep in
mind as they may differ in their sensitivity to errors and also suggest
a variety of quantum implementations. For instance, instead of the
matrix $A$ used above, defining
\(
B_{rs}={\left( -1\right) }^{{\left| 
r\cap s\right| }}\) gives 
\begin{equation}\vcenter{\halign{\strut\hfil#\hfil&#\hfil\cr 
$\displaystyle{B={\left( 
\matrix{B^{\prime}&B^{\prime}\cr 
B^{\prime}&-B^{\prime}\cr 
}\right) }}$\cr 
}}\end{equation}
which is unitary when normalized by dividing each element by $2^{N/2}$.
Thus \(
{\bf z}=B{\bf x}\) also has a simple recursive evaluation. This can be
used as the basis for computing the \(
V_{rjk}\) by noting that \eq{Vrj0} can be replaced with \(
\sum _{k=0}^{j}{\left( -1\right) }^{%
k}V_{rjk}=z_{r}\). Furthermore, instead of building up the values from
subsets with \eq{V relation}, an alternate relation is
\begin{equation}\vcenter{\halign{\strut\hfil#\hfil&#\hfil\cr 
$\displaystyle{\sum _{r^{\prime}}B_{rr^{%
\prime}}V_{r^{\prime}jk}=\sum 
_{l=0}^{j}V_{rjl}n_{lk{\left| 
r\right| }j}}$\cr 
}}\end{equation}
where the sum over \(
r^{\prime}\) includes {\em all} sets and
\(
n_{lk\rho \sigma }=\sum _{m=0}^{%
\rho }{\left( -1\right) }^{%
m}n_{lk\rho \sigma ;m}\) with
\begin{equation}\vcenter{\halign{\strut\hfil#\hfil&#\hfil\cr 
$\displaystyle{n_{lk\rho \sigma ;m}=2^{N-\rho -\sigma +l} \sum _{x=0}
^{l}{ 
{\rho -l} \choose {m-x} }
{ {l} \choose {x} }
{{\sigma -l} \choose {k-x} }
}$\cr 
}}\end{equation}
counting the number of sets \(
r^{\prime}\) with overlap $k$
with $s$ and
$m$ with
$r$, given that
$s$ and
$r$ have overlap
$l$ and sizes $\sigma$ and $\rho$
respectively. These relations involve somewhat more computational
operations on a classical machine than the method described above.
However, by avoiding use of the nonunitary mapping between each set and
its subsets, it may provide a method whereby quantum computers could
exploit the simple structure of the mapping.

\section{Quantum Search Behavior}
The behavior of this search algorithm was examined through a
classical simulation. While these results are limited to
small search problems, they nevertheless give an indication of how
this algorithm can dramatically increase the probability to find
solutions compared to the direct use of quantum parallelism. As a check
on the numerical errors, the norm in the final state was 1 to within
\(
10^{-10}\).

\figtwocol{x:example}{
\epsfig{file=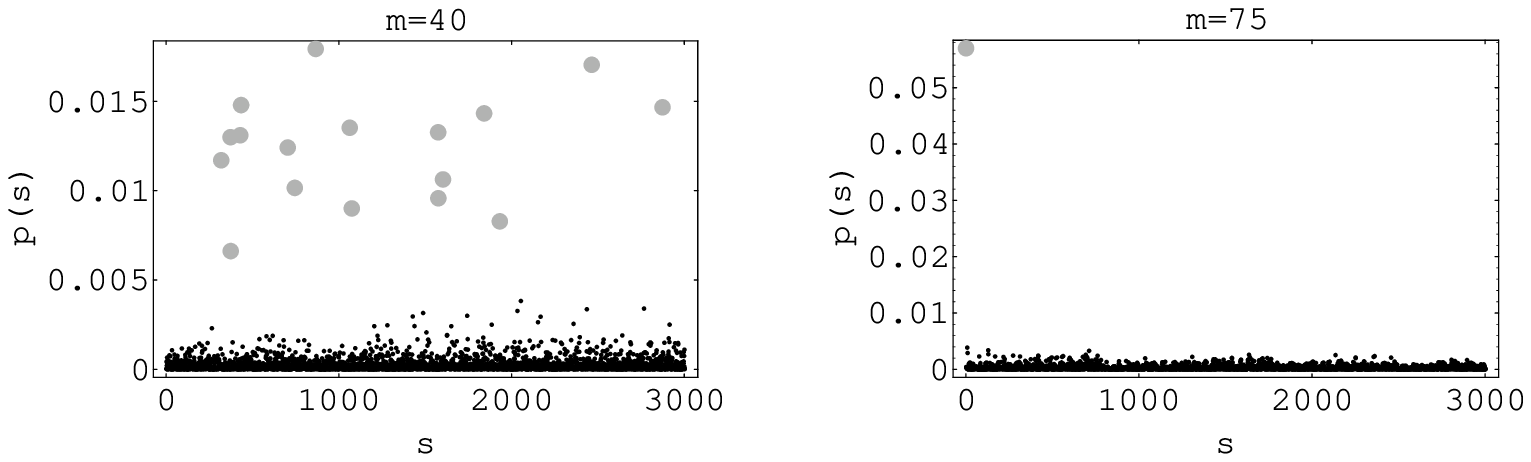}
}{Probability in sets at the solution level for two instances of 
unstructured problems with $N=15$. On the left, the problem has 
$m=40$, with 17 solutions and \(P_{\rm soln}=0.21\), on the right, 
$m=75$, with a single solution and \(P_{\rm soln}=0.057\). 
The solutions are drawn as large gray points. 
The remaining sets are small black points.}

\figtwocol{x:ratio scaling}{
\epsfig{file=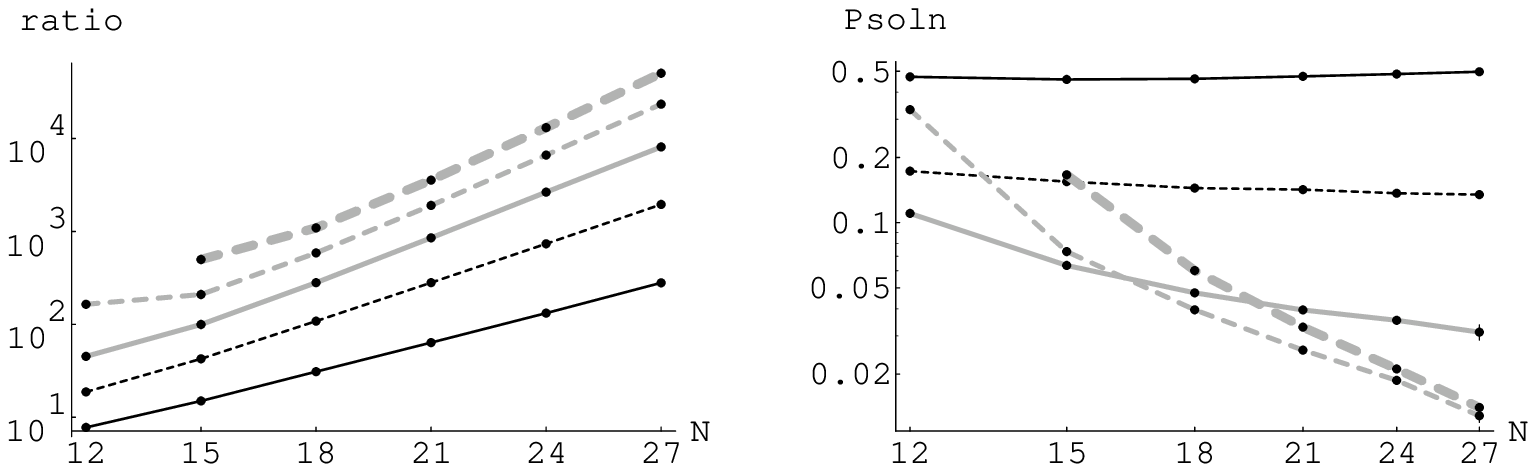}
}{Scaling of the algorithm. The left plot shows the ratio of 
probability to find a solution, \(P_{\rm soln}\), with the algorithm 
to that from random selection from among the solution level sets, for 
\(L=N/3\) with prespecified solution for \(\beta =m/N\) of 2 (solid), 
3 (dashed), 4 (gray), 5 (dashed gray) and 6 (thick dashed gray). 
This is shown on a log-scale. The right plot shows the scaling of \(P_{\rm soln}\) 
on a log-log plot. Each point is the average of 1000 problem instances 
(except only 100 were used for $N=27$). The points include error bars 
indicating the standard error of estimates of the mean, which in most cases 
are smaller than the size of the plotted points.}

As an example of the average behavior we consider a simple class of
unstructured problems corresponding to graph coloring with 3 colors. As
described above, a graph with $n$ nodes
has \(
N=3n\) variable-value pairs, and solutions consist of sets
of \(
L=n=N/3\) of these items. The constraints in a graph coloring problem
involve two items at a time. Thus to randomly generate an unstructured
version of such a problem, we select
$m$ distinct sets, each with two items,
to be the nogoods specified by the constraints. There are
\(
{ N \choose 2 }
\) such sets to choose from. This random selection
of problems ignores the detailed structure of the constraints for graph
coloring, but gives a wider range of possible
$m$ values for the small problems
considered here.

Since the
quantum algorithm is incomplete, i.e., it can find a solution but never
determine that no solution exists, we consider only problems with a
solution. Soluble problems are rare when there are many nogoods, so for
simplicity we use a prespecified solution, i.e., before selecting the
nogoods, a particular set of size $L$
is selected to be a solution. Then, the selection of
$m$ nogoods is only from among those
that are not subsets of this specified solution. This guarantees the
problem has at least one solution. Qualitatively similar results are
obtained by full random selection and then testing, via a complete
classical search, that the problem has a solution.


\fig{x:example} shows how the algorithm
enhances the probability of finding a solution for two instances: one
with few constraints and the other with many constraints. The sets are
ordered according to the integer whose binary representation corresponds
to including the items in the set~\cite{nijenhuis78}. In this case there are
\(
{ 15 \choose 5 }
=3003\) sets at the solution level, so a random selection
would give a probability of about 0.0003 to each set, much less than
given to solutions by this algorithm. Thus the various contributions to
nonsolutions tend to cancel out among the many paths through the
lattice.

\subsection{Scaling}


\fig{x:example} shows the improvement
over generate-and-test for two cases, but how well does the algorithm do
over the ensemble of problems, and how does this behavior scale with
increasing problem size? An appropriate choice of the scaling and method
of generating problem instances is necessary for a study of average
behavior so as to include a significant number of hard instances.
In this respect, an interesting scaling regime is when the
number of nogoods at level 2 grows linearly with the size of the problem
$N$, so we define \(
\beta \equiv m/N\). This corresponds to graph coloring where the number
of edges is proportional to the number of nodes, which has a high
concentration of hard search cases~\cite{cheeseman92}. 

For these problems, two views of the
scaling behavior are shown in  \fig{x:ratio scaling}. First, the linear 
growth on the log plot indicates this
algorithm improves exponentially compared to the direct use of quantum
parallelism. This is another indication of the effectiveness of this
simple algorithm at concentrating amplitude into solutions. The second
plot in the figure shows the overall probability to find a solution, on
a {\em log-log} plot, where linear behavior corresponds to a power law. Here
the problem sizes feasible for classical simulation are too small to see
the asymptotic behavior clearly. Nevertheless, it appears to do quite
well for problems with few constraints. And for the remaining cases, a
fit to a power law is closer than an exponential. We conclude that this
algorithm improves greatly on direct use of quantum parallelism, but it
is unclear whether that is enough to give polynomial rather than
exponential decrease of $P_{\rm soln}$, on average.

\subsection{Phase Transition}

Another indication of the usefulness of this quantum search
algorithm is its behavior as the number of constraints are changed for
problems of a given size. This is shown in \fig{x:transition}.
Significantly, the figure shows this search algorithm also
exhibits the transition behavior that was described above as occurring
for many classical searches~\cite{cheeseman92}, i.e.,
hard instances are concentrated near a change from underconstrained to
overconstrained problems.
Furthermore, the expected number of trials needed to find a solution is 
essentially
constant when there are relatively few constraints.
This appears to correspond to a second phase transition predicted to 
occur among weakly constrained problems~\cite{hogg94b}.
Thus the algorithm is using interference of paths to exploit
problem structure in the same manner as sophisticated classical search
methods are observed to do.

\figonecol{x:transition}{
\epsfig{file=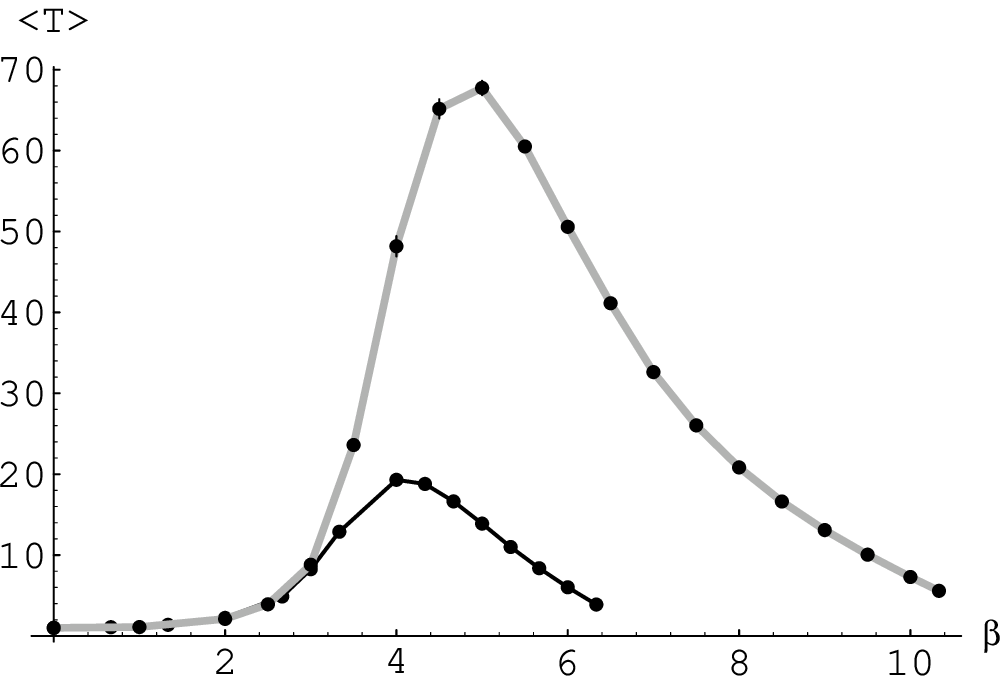,width=3.5in}
}{Average number of trials to find a solution, 
\({\left\langle T\right\rangle }=1/P_{\rm soln}\) as a function of $\beta$ 
for $N=15$ and 24 (black and gray curves, respectively). 
Each point is the average of 1000 problem instances, and 
includes error bars indicating the standard error of the mean, 
which in most cases are smaller than the size of the points.}

\figtwocol{x:tradeoff}{
\epsfig{file=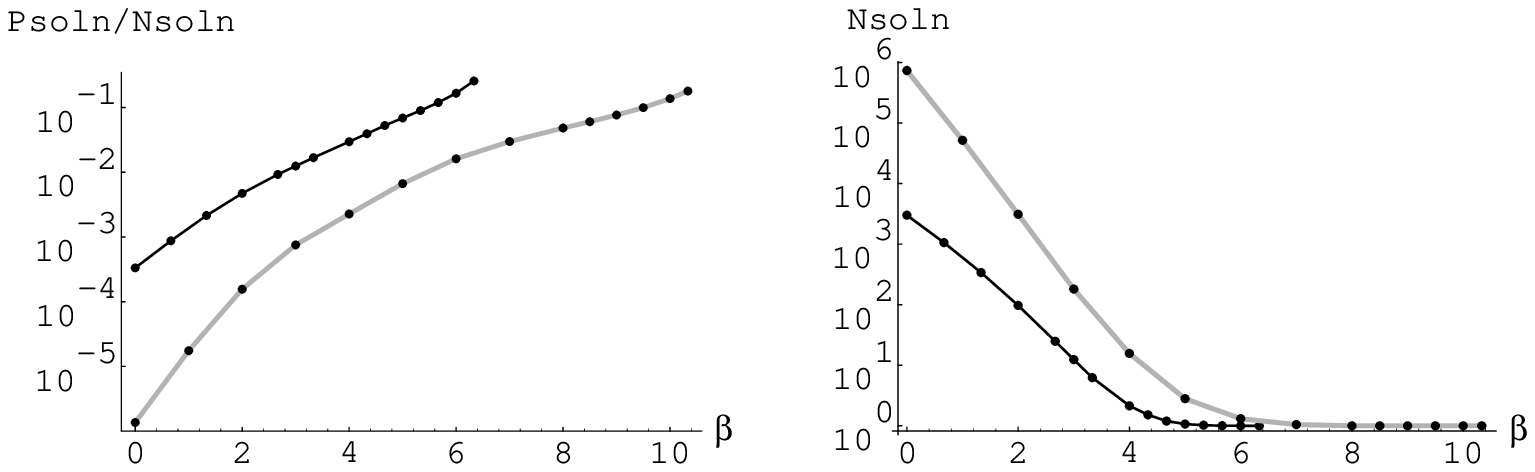}
}{The underlying trade-off giving the transition behavior for 
$N=15$ and 24 (black and gray curves, respectively). 
The left plot shows the average probability in each solution set, 
\(P_{\rm soln}/N_{\rm soln}\), as a function of $\beta$. 
The right plot shows the average number of solutions \(N_{\rm soln}\). 
Both plots are on a log scale. Each point is the average of 1000 problem instances, 
and includes error bars indicating the standard error of the mean, 
which are smaller than the size of the points.}

This can be understood from
\fig{x:tradeoff}, showing how the average
value for \(
p{\left( s\right) }\) among solutions compares with the expected number of
solutions \(
N_{\rm soln}\) for problems with differing numbers of constraints.
The average probability per solution grows as constraints are added. 
Thus the simple method of inverting phases at each nogood becomes
increasingly effective as more nogoods are added to the lattice.
However, this growth is initially overwhelmed by the even faster
decrease in the number of solutions, leading to a smaller chance to find
a solution, i.e., harder problems. Eventually the number of solutions
stops decreasing so rapidly because the constraints have already
eliminated most solutions. At this point the continued improvement in
concentration of amplitude into the remaining solutions dominates, so problems get
easier. Using such a figure to display the concentration
of amplitude into solutions can also be useful for viewing the behavior
of different choices for the phases introduced at inconsistent sets in
the lattice, perhaps in conjunction with the use of heuristics.

Similar behaviors have also been seen for other problem ensembles~\cite{hogg95d}. 
These include
cases corresponding to binary CSPs with two values per variable, i.e.,
\(
L=N/2\), and the NP-complete 3SAT problem where the
constraint nogoods are sets of size 3. For these types of problems,
slightly different ensembles can be produced through changes in the way
problem instances are generated. Although all these
empirical observations are limited to small problem sizes, they do
suggest the search method applies generally to a wide range of problem
ensembles.

\section{ Conclusion}
The lattice structure provides a general framework for applying
quantum computers to search problems. It has the advantage of an a
priori specification of the required computational steps and provides
many opportunities for using interference among the paths through the
lattice to each set at the solution level. While unlikely to work as
well as special purpose algorithms for particular search problems, this
framework may provide a basis for high performance search, on average,
over a wide range of problem types.

One important feature of this search framework is the ability to
incorporate additional knowledge about the particular problem structure
or other search heuristics. This is readily included as a modification
to the choice of phases because any such choice is guaranteed to be a
unitary operation and can operate independently on each state. Changes
to the mapping from one level to the next are more complicated due to
the requirement to maintain unitarity (as well as computational
simplicity). Another way to incorporate heuristics is by
changing the initial conditions. In the method reported here, initially
all amplitude is in small consistent sets. The hope then is that this
concentration in consistent sets is maintained as amplitude is moved up
through the lattice. Instead, one could start with amplitude spread
among all sets and try to arrange for it to be concentrated into
consistent sets only by the time it reaches the solution level. In some
cases, this is observed to increase the concentration into
solutions.

As a possible extension to this algorithm, it would be interesting
to see whether the nonsolution sets with relatively high probability
could be useful also, e.g., as starting points for a local repair type
of search~\cite{minton92}. If so, the benefits of
this algorithm would be greater than indicated by its direct ability to
produce solution sets. This may also suggest similar algorithms for the
related optimization problems where the task is to find the best
solution according to some metric, not just one consistent with the
problem constraints.

There remain a number of important questions. First is the issue of
implementation of the map from one level of the lattice to the next in
terms of more elementary operations that are physically realizable. This
could lead to the construction of special purpose search devices for the
set manipulations used in the problem-independent mapping between levels
of the lattice. Second, how are the results degraded by errors and
decoherence, the major difficulties for the construction of quantum
computers~\cite{landauer94a}? While there are some
quantum approaches to error control~\cite{berthiaume94,shor95} and 
studies of decoherence in the
context of factoring~\cite{chuang95} it remains to
be seen how these problems affect the framework presented here. Third,
it would be useful to have a theory for asymptotic behavior for large $N$, 
even if only
approximately in the spirit of mean-field theories of physics. This
would give a better indication of the scaling behavior than the
classical simulations, necessarily limited to small cases, and may also
suggest better phase choices. Considering these questions may suggest
simple modifications to the quantum map to improve its robustness and
scaling.
There thus remain many options to explore for using the deep structure of
combinatorial search problems as the basis for general quantum search
methods.

\section*{Acknowledgements}

I have benefited from discussions with J. Gilbert, J. Lamping and S. Vavasis.


\end{document}